# All-Spin Logic Device with inbuilt Non-Reciprocity

Srikant Srinivasan[1,2*], Angik Sarkar[1,2*], Behtash Behin-Aein[1,2*], and Supriyo Datta[1,2], *Fellow, IEEE*

[1]School of Electrical and Computer Engineering, Purdue University, W. Lafayette, IN 47907 USA
[2]NSF Network for Computational Nanotechnology (NCN), W. Lafayette, IN 47907 USA
*These authors contributed equally to this work.

**The need for low power alternatives to digital electronic circuits has led to increasing interest in logic devices where information is stored in nanomagnets. This includes both nanomagnetic logic (NML) where information is communicated through magnetic fields of nanomagnets and all-spin logic (ASL) where information is communicated through spin currents. A key feature needed for logic implementation is non-reciprocity, whereby the output is switched according to the input but not the other way around, thus providing directed information transfer. The objective of this paper is to draw attention to possible ASL-based schemes that utilize the physics of spin-torque to build in non-reciprocity similar to transistors that could allow logic implementation without the need for special clocking schemes. We use an experimentally benchmarked coupled spin-transport/ magnetization-dynamics model to show that a suitably engineered single ASL unit indeed switches in a non-reciprocal manner. We then present heuristic arguments explaining the origin of this directed information transfer. Finally we present simulations showing that individual ASL devices with inbuilt directionality can be cascaded to construct circuits.**

*Index Terms* — All Spin Logic, Magnetization Dynamics, Spin Transport, Spin Circuits, Spin Transfer Torques, Spintronic Logic Devices, Unidirectional Network.

## I. INTRODUCTION

DIGITAL electronic circuits store information in the form of capacitor charges that are manipulated using transistor-based switches. The need to find low power [1] alternatives has led to increasing interest in alternative schemes that store information in nanomagnets. This includes both nano-magnetic logic (NML) [2], [3] where the magnets communicate through their magnetic fields and all-spin logic (ASL) [4] where communication is through spin currents.

A key feature needed for logic implementation is non-reciprocity, whereby the output is switched according to the input but not the other way around. In NML this non-reciprocity has been achieved through Bennett clocking [2],[3],[5],[6], which requires additional clocking circuitry as compared to standard transistor-based logic. The objective of this paper is to draw attention to possible ASL-based schemes that utilize the established physics of spin-torque to provide an inbuilt non-reciprocity similar to transistors, thereby allowing logic implementation without the need for special clocking schemes.

*ASL design and operation*

First let us state clearly the operation of the proposed device and how introducing non-reciprocity can make it function in a deterministic manner. A generic ASL device is shown in Fig. 1(a), consisting of identical input ($FM_1$) and output ($FM_2$) ferromagnets connected to a common voltage source $V_{SS}$ that drives a charge current across the ferromagnets to a ground contact through a common non-magnetic channel. Although both magnets are connected to the same voltage, significant spin current ($\vec{I_S}$) is exchanged between them as will be shown in subsequent sections.

The black regions under the magnets in Fig. 1(a) represent isolation layers in the conducting region of the channel. Such design ensures that communication between the two magnets is limited to the channel segment between the two of them. Therefore, each magnet can interact independently with different magnets on either side. Specifically each magnet can have a "talking" side that behaves as an input to the following

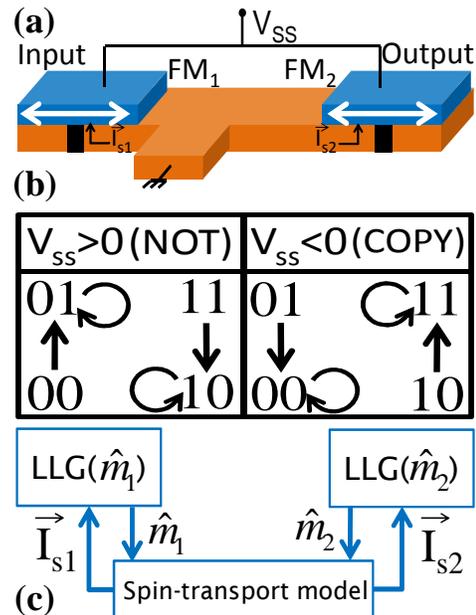

Fig.1. The basic operation of a generic ASL device. (a) A suggested layout showing two identical Ferromagnets, one ($FM_1$) acting as input and the other ($FM_2$) acting as output. (b) This table illustrates the basic operation of the device assuming $|V_{SS}|$ to be larger than the minimum value needed for switching. These state diagrams describe the transitions of $FM_1$ (first bit) and $FM_2$ (second bit) for NOT and COPY operations. (c) An illustration of the coupled spin-transport / magneto-dynamics model. Here, $\vec{I_s}$ is spin current and $\hat{m}$ is the magnetization direction.

Corresponding author: S.Datta (e-mail: datta@purdue.edu ).

stage and a "listening" side that behaves as an output to the previous stage thus allowing for cascaded ASL structures. Fig. 1(b) shows the state diagram for two interacting magnets ($FM_1$ and $FM_2$) connected by one such channel segment and can be understood as follows:

Ordinarily, with the supply voltage $V_{SS} = 0$, the two magnets in Fig.1(a) can exist in any of four possible states that we denote as 00, 01, 10, 11 ($FM_1$: first bit, $FM_2$: second bit). However, we will show that as $V_{SS}$ is increased beyond a positive threshold value the only stable states are those for which the two magnets are *anti-parallel*, namely 01 and 10. On the other hand, if $V_{SS}$ is negative then beyond a certain threshold value, the only stable states are those for which the two magnets are *parallel*, namely 00 and 11. These stable states are the ones shown in Fig. 1(b) with arrows pointing back to themselves.

The key point regarding the inbuilt non-reciprocity of this device can be appreciated by considering what happens if we start from a state, say 00 that is rendered unstable when the supply voltage $V_{SS}$ is positive and exceeds a critical value. In principle it could make a transition to either of the two stable states 01 or 10. However, we will show that it should be possible to engineer the device so that it will transition predictably to 01 and not to 10 as indicated in Fig. 1(b). In other words, the transition arrows in the table are always "vertical" and never "horizontal", since it is always $FM_2$ that changes its magnetization appropriately to reach a stable state, while $FM_1$ never switches its magnetization.

What we wish to emphasize here is that the identical magnets $FM_1$ and $FM_2$ function like the input and the output respectively, with an input-output isolation that is characteristic of transistors. The operation with $V_{SS} > 0$ can be viewed as a NOT operation, while the operation with $V_{SS} < 0$ can be viewed as a COPY operation.

*Experimental realization of ASL*

In this paper we will establish the feasibility of the above claims regarding ASL operation using a coupled spin-transport / magnet-dynamics model (Fig.1c). This coupled model, the details of which are given in [9], describes existing experiments such as [7] quite well. These experiments demonstrate feasibility of switching a magnet in structures similar to Fig. 1(a) with voltages of the order of a few 10s of milli-volts. Such low-voltage operation is very attractive for switching applications. However, we would like to clearly state which features of ASL operation described earlier are experimentally established and which – at this time – are not.

First, in the present ASL structure $FM_1$ and $FM_2$ are both identical free layers, either of which could switch as easily as the other. By contrast, all existing experiments [7], [8] utilize a free layer and a fixed layer, such that only one magnet can switch. While having a fixed layer clearly ensures "non-reciprocity", it is not suitable for logic implementation since the output of one section must subsequently function as the input to another. This requires that both the magnets should be nominally identical.

What is it then that differentiates $FM_1$ from $FM_2$ in an ASL device, making them act differently as input and output respectively? We will discuss various possibilities in subsequent sections such as the one shown in Fig. 2(a). Here, simply having the ground terminal closer to one of the magnets makes it function as an input as supported by our simulations in Fig. 3.

Second is the application of the same supply voltage $V_{SS}$ to both the input and output magnets, which is attractive from a practical point of view since it is easier to apply the same voltage to closely-spaced structures and also minimizes the associated capacitive charges. Normally if we wish one magnet to turn another, it may not seem natural to apply the same voltage to the two magnets. Indeed, the experiment in [7] had the output magnet left floating as is common with non-local spin valve experiments [13]. However, it will be shown later that the final stable states of $FM_1$ and $FM_2$ being parallel or anti-parallel to each other is a feature peculiar to having the same supply voltage on both magnets. Additionally, it will also be shown in Appendix B that the operational regime with different supply voltages to the input and output magnets has its own interesting characteristics that could have potential applications.

*Outline*

As discussed earlier the key features distinguishing ASL operation (Fig. 1(b)) are:
- Its transition diagram allows only "vertical transitions" dictated by the input magnet.
- *Stable states (01, 10 for Vss>0 and 00, 11 for Vss <0).*

These two features are first established in Section II through detailed numerical simulations. Next, Section III provides an intuitive understanding of the ASL state diagram.

In section IV we show that ASL devices (each identical to the one in Fig. 2(a)) with inbuilt directionality can be cascaded to construct circuits. Our simulations suggest that three identical nanomagnets (Fig. 5) linked by spin transport channels can form a chain of inverters with a directed transfer of information (magnetization) from one magnet to the next: 1→2→3. This inbuilt directivity also allows the last magnet in the chain to drive the first one, when connected, to function like a classic ring oscillator (Fig. 6) well known in CMOS circuits.

## II. Establishing the ASL state diagram

In this section we will present numerical evidence supporting the state diagram in Fig. 1(b) using an experimentally benchmarked model [9]. Fig. 2(a) shows one implementable scheme for introducing non-reciprocity in the ASL device. The difference from Fig. 1(a) is that here the lead to ground is shifted to the channel region lying below $FM_1$ thus making it the input. $FM_1$ effectively shields $FM_2$ from communicating with the ground terminal, thereby reducing the charge current and consequently the spin current ($\vec{I_s}$) injected by $FM_2$ as compared to $FM_1$. As a result the torque exerted on $FM_2$ given by $\vec{I_s^\perp} = \hat{m}_2 \times (\vec{I}_{s2} \times \hat{m}_2)$ is greater than the torque exerted on $FM_1$. This effect is captured by the vicinity of FM$_2$ compared to the $\hat{x}$ spin voltage in the vicinity of FM$_1$. The ability of the input magnet to shield the output magnet from the ground terminal depends on the ratio of the magnet contact length ($L_C$) to the transfer length of the contact which is of the order: $L_{rg} = 1/\sqrt{\frac{g\rho}{WA}}$ where g=contact conductance, W=width of the channel, $\rho$=channel resistivity; A=cross-sectional area of the channel. The ratio of spin-torques $I_{S2}^\perp / I_{S1}^\perp$ follows from $V_Z/V_X$ as is evident by comparing Figs. 2 (c) and (d). Fig. 2(c) was simulated for the particular case of $L_C \sim L_{rg}$ and the corresponding point can be identified in Fig. 2(d).

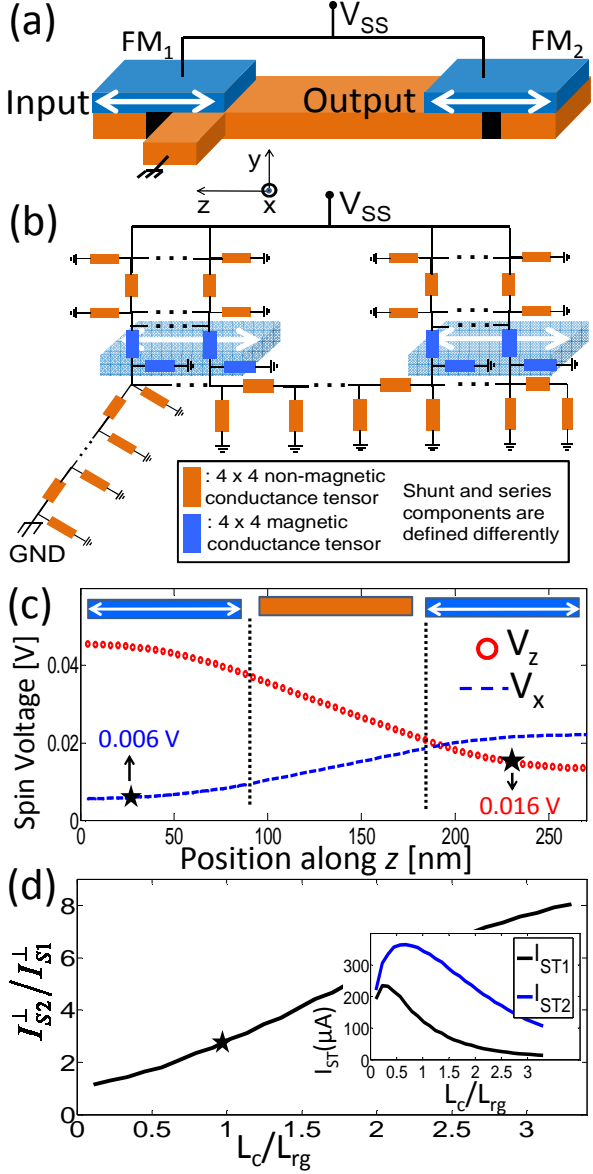

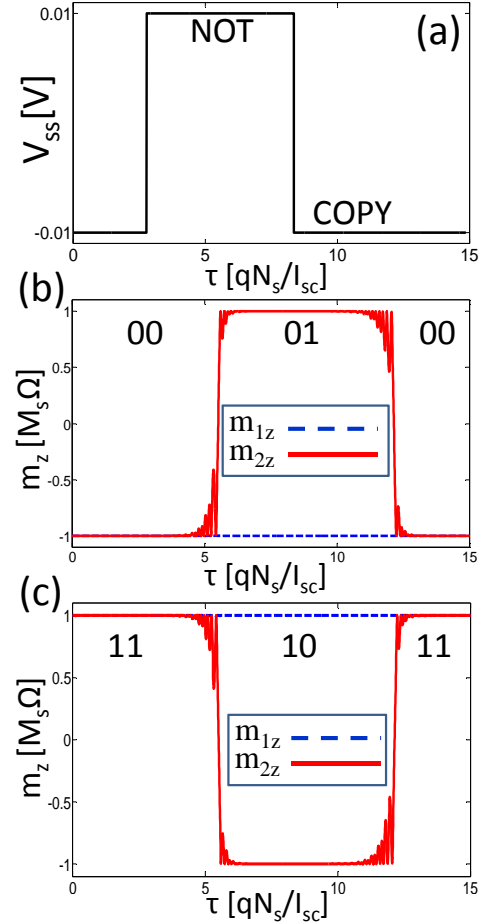

Fig.2 A cascadable ASL design that achieves non-reciprocity through asymmetric placement of the ground terminal (a) The position of the ground terminal is closer to the input rather than output. (Not drawn to scale) (b) The distributed conductance network used to model the device. Each conductance element is a 4 by 4 tensor. The shunt and series conductance elements are defined differently. (c) Shows the in $z$ and $x$ components of spin voltage in the channel region assuming input magnet is along $z$ and output magnet is along $x$. A higher spin voltage underneath the output magnet as compared to the input magnet results in higher spin torque current getting exerted on output as compared to input. (d) The ratio of torques exerted on FM$_1$ (input) and FM$_2$ (output) magnets as a function of contact length normalized to the '$rg$' length.

distributed version (Fig. 2(b)) of the conductance model for spin transport introduced in [9] which accounts for the spatial variation of quasi-Fermi levels in the channel for charge as well as each spin component. We use the same parameters as in the experiment [7] to present our results for identical magnets. As an illustrative example, Figs. 2 (c) and (d) show the numerical results for the specific case when FM$_1$ is along $\hat{z}$ and FM$_2$ is along $\hat{x}$. Fig. 2(c) shows how a higher injection by FM$_1$ results in a larger $\hat{z}$ spin voltage distribution in the

Fig.3 Coupled spin-transport / LLG simulations describing the NOT and COPY operations in an ASL device with inbuilt non-reciprocity. (a) The voltage pulse applied to the structure of Fig.2a. (b) Both magnets are initialized at 00. With $V_{SS}<0$, the parallel states are stable and 00 state is retained. But when the voltage is changed to $V_{SS}>0$, the output becomes the NOT of input and anti-parallel state (01) is reached which is now stable. System reaches 00 again when voltage is changed back to positive values and is stable again. (c) Similar simulations as in part (b) but with the two magnets initialized at 11 state.

Fig. 3 shows a typical simulation result for such a structure from our coupled model (Fig.1c) of spin-transport and dynamics of magnetization ($\widehat{m}$) described by the Landau-Lifshitz-Gilbert (LLG) equation:

$$\frac{d\widehat{m}}{dt} = -|\gamma|\widehat{m} \times \vec{H}_{\text{int}} + \alpha\widehat{m} \times \frac{d\widehat{m}}{dt} + \frac{\vec{I}_s^\perp}{qN_s} \qquad (1)$$

Where $\gamma$ is the gyromagnetic ratio, $\alpha$ is the Gilbert damping coefficient, $q$ is the charge of an electron. $N_s$ is the net number of Bohr magnetons in a nanomagnet given by $N_s = M_s \Omega / \mu_B$ ($M_s$: saturation magnetization, $\Omega$: volume and $\mu_B$: Bohr magneton). $\vec{H}_{int}$ describes the internal fields of a magnet. Note that each magnet is described by its separate LLG equation.
The solid curves in Figs. 3(b, c) show the state of the output magnet and the dashed curves show the state of the input magnet as a function of the dimensionless time τ in units of $qN_s/I_{sc}$. $I_{sc}$ is the critical spin-current [10] needed for switching given by:

$$I_{sc} = \frac{2q}{\hbar}(2\alpha E_b)(1 + H_d / 2H_K)$$

$\hbar$ is the reduced Plank's constant. $E_b$ is the anisotropy energy barrier. $H_d$ is the demagnetizing field and $H_K$ is the uniaxial anisotropy field. Note that the absence of $H_d$ in perpendicular magnetic anisotropy magnets [18] would reduce the critical spin current. Such issues are discussed in [9].

These figures show that when $V_{SS}<0$, the initial parallel states (00 and 11) are retained and are stable. Once $V_{SS}>0$, the output becomes the NOT of the input resulting in anti-parallel states (01 and 10) which under positive voltages are stable states. As $V_{SS}$ is changed back to negative values, the output once again becomes the copy of the input and the resulting parallel states are stable. Our coupled model effectively establishes the state diagram in Fig. 1(b).

III. "UNDERSTANDING" THE STATE DIAGRAM

Now that we have presented numerical evidence supporting the transition diagram (Fig. 1(b)), let us try to understand the underlying physics.

*A) How the stable states are determined*

As indicated in Fig.1b, with a negative voltage $V_{SS}$, the parallel states 00,11 are stable, while with a positive $V_{SS}$, the anti-parallel states 01,10 are the stable ones. To understand this, we first note that in general, the spin-currents entering the two magnets can be written as [17]:

$$\vec{I}_{s1} = A_1 \hat{m}_1 + B_1 \hat{m}_2 + C_1(\hat{m}_1 \times \hat{m}_2) \quad (2a)$$
$$\vec{I}_{s2} = A_2 \hat{m}_2 + B_2 \hat{m}_1 + C_2(\hat{m}_2 \times \hat{m}_1) \quad (2b)$$

so that the spin-torques (s.t.) due to these spin-currents can be written as (Note: $c \equiv \hat{m}_1 \cdot \hat{m}_2$)

$$qN_s \frac{d\hat{m}_1}{dt}\bigg|_{s.t.} = \vec{I}_{s1} - (\vec{I}_{s1} \cdot \hat{m}_1)\hat{m}_1$$
$$= -cB_1\hat{m}_1 + B_1\hat{m}_2 + C_1(\hat{m}_1 \times \hat{m}_2) \quad (3a)$$
$$qN_s \frac{d\hat{m}_2}{dt}\bigg|_{s.t.} = \vec{I}_{s2} - (\vec{I}_{s2} \cdot \hat{m}_2)\hat{m}_2$$
$$= -cB_2\hat{m}_2 + B_2\hat{m}_1 + C_2(\hat{m}_2 \times \hat{m}_1) \quad (3b)$$

These spin-torque currents are only a part of the terms that enter the right-hand side of the LLG equation. But suppose we ignore the other terms involving the internal fields and use Eqs. 3 (a,b) to write

$$\Rightarrow \frac{d}{dt}c = \frac{d\hat{m}_1}{dt} \cdot \hat{m}_2 + \frac{d\hat{m}_2}{dt} \cdot \hat{m}_1 = (B_1 + B_2)(1-c^2) \quad (4)$$

This equation of course does not include the effective internal fields and as such should not be taken too seriously. But it is easy in that it does explain the stable states correctly.

Clearly from Eq. (4) the fixed points where dc/dt =0, are given by $1-c^2 = 0$, that is, $c = +1$ (parallel magnets) and $c = -1$ (anti-parallel magnets). Of these two, however, only one is stable, the one for which $(B_1 + B_2)c > 0$, so that the "Jacobian" for Eq. (4) is negative.

For a positive $(B_1 + B_2)$ this means that the stable configuration is $c = +1$ corresponding to parallel while for negative $(B_1 + B_2)$ the stable configuration is $c = -1$ corresponding to anti-parallel. This would explain the stable states in Fig.1b, if the sign of $(B_1 + B_2)$ is opposite to that of $V_{SS}$. This is indeed true and can be understood as follows. A negative $V_{SS}$ applied to a ferromagnet results in the channel being populated preferentially by spins in the direction of the injecting ferromagnet [15] giving a positive value for B, while a positive $V_{SS}$ populates the channel preferentially with spins in the direction opposite to the ferromagnet, resulting in a negative value for B.

*B) Why transitions are vertical*

It can be shown that initially even if both magnets start to switch in Figs. 3(b,c) the output magnet gets going faster and soon the input magnet goes back to its initial state, since the overall objective of both magnets is the same: they both want to be anti-parallel for $V_{SS} > 0$, and parallel for $V_{SS} < 0$.

Why does one magnet get going faster? Essentially because the other magnet being closer to the ground terminal is more efficient in generating the spin currents needed to switch it. This difference can be quantified by considering a gedanken experiment. Suppose using the device in Fig. 2(a), we first hold $FM_1$ fixed and only allow $FM_2$ to switch. At low voltages, the spin current injected by $FM_1$ does not provide sufficient torque to affect $FM_2$. As the voltage is ramped up (Fig. 4), $FM_2$ gets switched at some voltage $V_{c2}^+$. Next suppose the experiment is reversed: $FM_2$ is held fixed and $FM_1$ is allowed to switch which occurs at a different critical voltage $V_{c1}^+$. Similar experiments can be repeated for $V_{SS}<0$ to obtain the corresponding critical voltages. For a perfectly symmetric ASL, $V_{c1}^+=V_{c2}^+$. The difference between $V_{c1}$ and $V_{c2}$ arises from any asymmetry, intentional or unintentional, in the structure.

For example, Fig. 4 was generated for the situation where $FM_1$ and $FM_2$ differ only in the placement of the ground lead as illustrated in Section II (A). This asymmetry dictates that a higher voltage is required to generate enough spin current to switch $FM_1$. In essence, the window $V_{c1}$-$V_{c2}$ is a direct measure of the non-reciprocity in the system.

$V_{c1}$ and $V_{c2}$ delineate the range of supply voltages into 3 distinct regimes. When $|V_{SS}|<V_{c2}$ no magnet has enough spin current to switch. So, all four states 00,10,11,01 are stable. As the voltage is further increased $V_{c2}<|V_{SS}|<V_{c1}$ only $FM_2$ has enough current to switch and the magnets finally stabilize in anti-parallel (parallel) for positive (negative) $V_{SS}$. The stability in this regime is expected since the operation is in a window where only one magnet can switch.

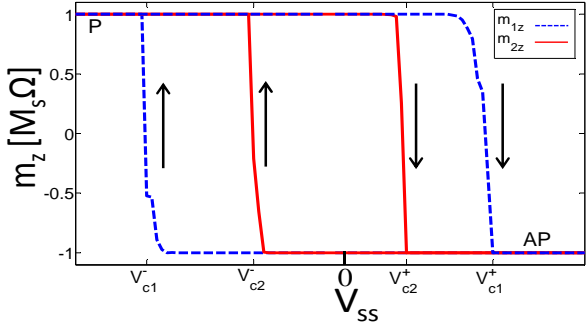

Fig 4. This figure shows different switching voltages for *identical* input and output magnets. The anti-parallel (AP) states are stable for positive $V_{SS}$ while parallel (P) states are stable for negative $V_{SS}$. $FM_1$ and $FM_2$ differ only in the placement of the ground lead as illustrated in Section II (A).

When $V_{c2}<V_{c1}<|V_{SS}|$, there is sufficient torque acting on both $FM_1$ and $FM_2$ to switch either of them. However, it is seen that it is only $FM_2$ that switches. The reason is that the induced asymmetry ensures that the torques on $FM_2$ is greater than the torque on $FM_1$ and the time that they would have taken to switch is different. Initially, both magnets start to switch. However, as soon as $FM_2$ switches, $FM_1$ relaxes to its original state which is the stable state of the two magnet system as discussed before. For perfect reciprocity the switching behavior is not predictable. In reality, an asymmetry will ensure which magnet will predictably switch and the challenge is to deliberately engineer an asymmetry large enough to overcome random variations and enforce the desired switching behavior. We will discuss additional schemes for engineering such asymmetry in the appendix.

## IV. CASCADING ASL DEVICES

Having established numerically and analytically in previous sections, the validity of ASL operation given by Fig. 1(b), we now show that multiple ASL devices with inbuilt directionality (such as in Fig. 2(a)) can be cascaded to construct circuits [16].

Let us first consider the simple example of 3 identical magnets connected in series as shown in Fig. 5. When a positive voltage is applied to this circuit, each magnet tries to invert the ones connected to it. However, the placement of the ground lead enforces a directed transfer of information (magnetization) in a specific order: 1→2→3. Consequently the circuit behaves as a chain of inverters. This is illustrated in Fig. 5(b) where initially both $FM_1$ as well as $FM_2$ start to switch. However, $FM_1$ dominates and it is $FM_2$ that gets inverted from logic '0' to '1'. Subsequently $FM_2$ inverts $FM_3$ and the system becomes stable.

It is important to note that such directed transfer of information could be attempted using either non-identical magnets (such as making them progressively smaller) or using additional clocking circuitry. However, additional clocking would introduce its own design complexity and power dissipation issues. Similarly, using non-identical magnets is not feasible in the context of integrated circuits with many devices. Also from an operational point of view the 'small' last magnet of the chain wouldn't be able to drive earlier stages thus preventing information transfer in a loop.

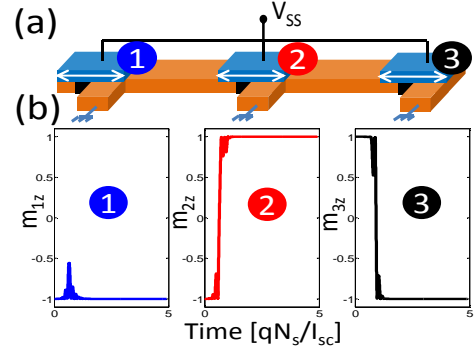

Fig. 5: A chain of inverters showing the cascading of ASL devices. (b) The three magnet system is initialized at 001. At time t=0, the voltage is turned on. Once the steady state is reached, the final state of the three magnet system becomes 010. Since directivity is present, magnet 1 dictates the state of magnet 2 and that in turn decides the state of magnet 3.

The power of inbuilt directivity using identical magnets is clearly illustrated by the example of a classic ring oscillator (Fig. 6) well known in CMOS circuits. It allows sufficient feedback from the last magnet in the chain of identical magnets to drive the first one without any clocking circuitry.

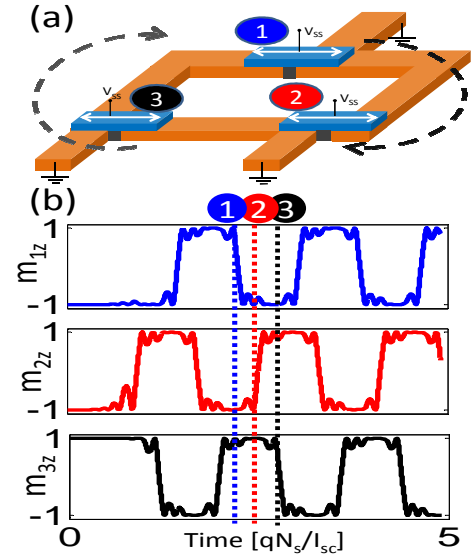

Fig 6. (a) A ring oscillator composed of 3 identical magnets in cascaded ASL devices. (b) This system will oscillate with a positive DC supply voltage applied to the magnets. The direction of information flow is 1→2→3. The process continues in a cyclic fashion. The switching edges – identified by the vertical dashed lines – for magnets 1, 2 and 3 show the step by step process of information transfer from one magnet to the next.

## V. CONCLUDING REMARKS

Electronic devices have traditionally been based on controlling the flow of charge. However, electrons carry both charge and "spin", the latter being responsible for magnetic phenomena. In the last ten years there have been significant advances in our ability to control the spin current in electronic devices and their interactions with nanomagnets. A recent proposal for All-Spin Logic (ASL) proposes to store information in nanomagnets which communicate using spin currents for communication. ASL is defined by: *Information:* Magnetization of magnet, *Communication:* Spin current, *Energy:* Power supply. This is similar to the charging of an output capacitor according to the information provided by the charge on an input capacitor in standard CMOS logic: *Information:* Charge on capacitor, *Communication:* Charge current, *Energy:* Power supply.

However, a key element in any approach for logic implementation is the input-output isolation that comes so easily in transistors that it often goes unnoticed. The purpose of this paper is to present a unique approach of achieving non-reciprocal operation with ASL devices based on spin-torque physics that could enable combinational logic similar to CMOS. Using a model for ASL devices that is based on established physics and is benchmarked against available experimental data, we present interesting predictions that we hope will inspire creative experimentation along these lines. The ring oscillator and chain of inverters shown in this paper are just two demonstrative examples of all-spin computation. The principles put forth in this paper can be used to build more complex circuits which we leave as future work.

## APPENDIX

### A) OTHER SCHEMES FOR NON-RECIPROCITY WITH COMMON VSS

The ASL state diagram (Fig. 1(b)) was established earlier using asymmetric ground layout. However, this could have been achieved through other structural asymmetries. To investigate other possibilities let us first quantify the non-reciprocity. The degree of non-reciprocity can be arrived at by defining a "spin-torque conductance" relating the spin-torque component of the current at each of the magnets in the ASL device to the externally applied voltage:

$$|\vec{I}_{s1}^\perp| \equiv |\hat{m}_1 \times (\vec{I}_{s1} \times \hat{m}_1)| = g_{s1} V_{SS}$$
$$|\vec{I}_{s2}^\perp| \equiv |\hat{m}_2 \times (\vec{I}_{s2} \times \hat{m}_2)| = g_{s2} V_{SS}$$

The magnet with the larger magnitude of $g_s$ functions as the output.

Fig. A1 shows the simplest conductance model to analyze the ASL structure (Fig. 1(a)) with the ground lead now in the centre of the channel. This model allows us to obtain an exact analytical expression that matches the distributed numerical model in Fig. 2(b) for near ballistic channels.

$$\frac{g_{s2}}{g_{s1}} \equiv \frac{|I_{S2}^\perp|}{|I_{S1}^\perp|} = \left(\frac{g_{\beta 2}}{g_{\beta 1}}\right) \frac{(2 g_{\alpha 2} \cos^2 \frac{\theta}{2})(g_{\beta 1} - g_{F1}) + g_{\alpha 1}(g_{0P} + g_{F2}) + g_{\alpha 2} g_{F1} + (g_{\alpha 1} - g_{\alpha 2}) g_{\beta 1}}{(2 g_{\alpha 1} \cos^2 \frac{\theta}{2})(g_{\beta 2} - g_{F2}) + g_{\alpha 2}(g_{0P} + g_{F1}) + g_{\alpha 1} g_{F2} + (g_{\alpha 2} - g_{\alpha 1}) g_{\beta 2}} \quad (A1)$$

Note that each of the conductances is a 4x4 matrix and relates 4x1 voltages and currents that include charge (*c*), and the three (*x,y,z*) spin components i.e. [c,z,x,y]$^T$.

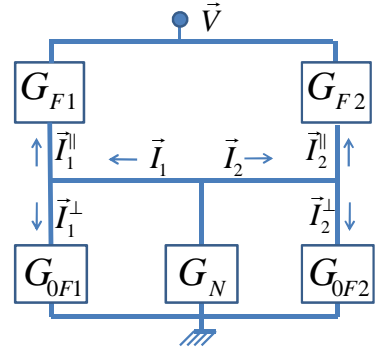

Fig. A1: Circuit representation of an ASL device for a one-point channel that is used to derive a relatively simple closed form equation (A1) describing devices with near ballistic channels. (Note that the simulations use a distributed conductance network as illustrated in Fig.2b)

For a Ferromagnet in the '*z*' direction, the interface conductances are given by $G_F$ and $G_{0F}$ as:

$$G_F = \begin{pmatrix} g_F & g_\alpha & 0 & 0 \\ g_\alpha & g_F & 0 & 0 \\ 0 & 0 & 0 & 0 \\ 0 & 0 & 0 & 0 \end{pmatrix}, \quad G_{0F} = \begin{pmatrix} 0 & 0 & 0 & 0 \\ 0 & 0 & 0 & 0 \\ 0 & 0 & g_\beta & -g_\gamma \\ 0 & 0 & g_\gamma & g_\beta \end{pmatrix}$$

where $g_F$ is the conductance of the FM/interface region, $g_\alpha = P g_F$ and P is the effective polarization of the FM interface. $g_\beta$ and $g_\gamma$ refer to the effective spin 'mixing' conductance of the interface as defined in [11] and describe the Slonczewski and Field-Like components of Spin Torque respectively. In the subsequent analysis, we set $g_\gamma = 0$ because in general the field-like term is found to be very small in all-metallic structures [14]. We assume the lead to the ground terminal to be unpolarized with the absence of any spin-orbit interaction effects, thereby, equally affecting all spin components. It can be described by a matrix ([9] and [12], [13] therein)

$$G_N = \begin{pmatrix} g_0 & 0 & 0 & 0 \\ 0 & g_{0P} & 0 & 0 \\ 0 & 0 & g_{0P} & 0 \\ 0 & 0 & 0 & g_{0P} \end{pmatrix} \quad \text{where } g_0 = \frac{A}{\rho L}; \\ g_{0P} = \frac{A}{\rho \lambda} \coth(\frac{L}{\lambda})$$

$\rho$, $\lambda$, L and A refer to the resistivity, spin-diffusion length, length and cross section of the lead respectively.

Solving the conductance model in Fig. A1 gives us an expression for the degree of non-reciprocity given by A1:

It is instructive to look at a simplified version of Eq. (6) for two perpendicular magnets ($\theta$ = pi/2), for which the expression simplifies to

$$\frac{|g_{S2}|}{|g_{S1}|} = \frac{g_{\alpha 1} g_{\beta 2} (g_{\beta 1} + g_{0P} + g_{F2})}{g_{\alpha 2} g_{\beta 1} (g_{\beta 2} + g_{0P} + g_{F1})} \quad (A2)$$

Eq. A2 suggests several possibilities for implementing non-reciprocity in a controlled way by engineering the various components of the device structure. Among other things the non-reciprocity depends on how well the input magnet can inject polarized spin current ($g_{\alpha 1}$) and how easily the output magnet can relax the non-collinear spins ($g_{\beta 2}$) and vice-versa.

Different schemes for introducing non-reciprocity include: (1) Insertion of a tunnel barrier at the interface of one side of the magnets (2) Defining different interface areas for the input and output (3) Novel techniques for designing different injection efficiencies at the input and output magnet. Such schemes could involve simultaneously altering the effective polarization as well as resistance of the interface and these effects would all be captured by Eq. (A1). These different schemes all need to be explored carefully. Our purpose here is simply to identify the relevant parameters, using a simple static analysis based on the spin-circuit model.

*B) Non-reciprocity using different voltages*

It is also possible to introduce non-reciprocity in a physically symmetric structure simply through applying different voltages at the two magnets. We show one such case in Fig. A2(a) where one of the magnets is connected to Vss and the other is grounded.

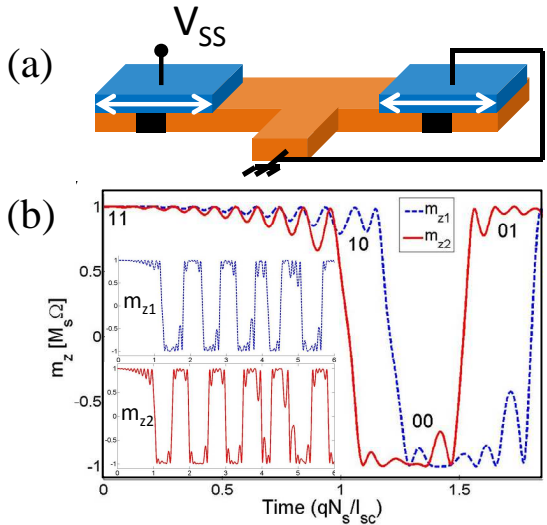

Fig. A2: An structurally symmetric ASL device with non-reciprocity induced by two different voltages applied to the input and output magnets. (a) The output magnet shorted to ground (b) The evolution of the two magnets in this case with application of $V_{SS}>0$. (inset) The magnets make astable periodic oscillations.

The non-reciprocity in this case is given by:

$$\frac{|g_{S2}|}{|g_{S1}|}=1+\frac{g_0/g_F}{1-\frac{2g_\alpha^2\cos^2\frac{\theta}{2}}{g_F(g_F+g_{0P}+g_\beta)}}$$

What happens in general is that the magnet with a supply voltage closer to that of the ground terminal ends up with a higher value of $g_S$. However, this system has no stable state and once the output has switched, the system continues to oscillate deterministically between all the states 00,01,11,10. The oscillations are primarily due to the fact that magnets try to enforce opposite configurations as one of them is injecting spins while the other is extracting spins. While these oscillations make it difficult to implement logic in this configuration, the periodic astable oscillations might have their interesting (Fig A2 (b) insets) applications.

ACKNOWLEDGMENT

This work was supported by the Institute for Nanoelectronics Discovery and Exploration (INDEX) under the Nanoelectronics Research Initiative (NRI) program as well as NSF Center for Science of Information (CSoI). The authors are grateful to Joerg Appenzeller and Sayeef Salahuddin for insightful discussions on possible experimental layouts. S.S. also thanks Kerem Y. Camsari for helpful discussions.